\documentclass[preprint,12pt]{elsarticle}
\usepackage{graphicx}
\usepackage{pgf}
\usepackage{amsmath}
\usepackage{amssymb}
\usepackage{mathrsfs} % para formato de letra
\usepackage{theorem}

\usepackage{pstricks}
\usepackage[utf8]{inputenc}

\usepackage[normalem]{ulem}

\journal{looking for}

\begin{document}

\begin{frontmatter}

%\runauthor{Antonio Sala}
%\begin{frontmatter}%elsevier
\title{Interlacing in Controllers Implementation: Frequency Analysis}
\author{Juli\'{a}n Salt}
%\thanks{J. Salt was Visiting Scholar at Department of Mechanical Engineering, Universiy of California at % Berkeley. USA }
%A. Sala is with the Systems Engineering Department, Universidad
%Politecnica de Valencia, Cno. Vera, s/n, E-46022 VALENCIA, SPAIN.\
%e-mail: asala@isa.upv,es, julian@isa.upv.es} }
\address{Systems Eng. and Control Dept., Instituto de Automatica e Informatica Industrial, Universitat
Polit\`{e}cnica de Valencia, Cno. Vera, s/n, E-46022 VALENCIA, SPAIN.%\\
%A. Sala is with the Systems Engineering Department, Universitat
%Politecnica de Valencia, Cno. Vera, s/n, E-46022 VALENCIA, SPAIN.
\\
e-mail: julian@isa.upv.es}
%\thanks[Someone]{Partially supported by somebody.}
%\date{}
%\maketitle

\begin{abstract}  

The main goal of this contribution is to explain how to use interlacing techniques for LTI controllers implementation and analyze different structures in this environment. These considerations lead to an important computation saving in constrained resource environments. It has been also introduced new procedures for obtaining the blocks related to different real and complex controller's poles. The resultant time-varying system is modeled using proper discrete lifting techniques and a new and efficient dual-rate frequency response computation allows to determine the characteristics of the control loop with interlaced controller. Examples illustrate the theoretical proposals.
\end{abstract}

%\end{frontmatter}
%\title{Frequency-response of dual-rate systems}
%\maketitle
\end{frontmatter}

\section{Motivation}

\begin{figure} \centering
\includegraphics[height=9cm, width=12cm]{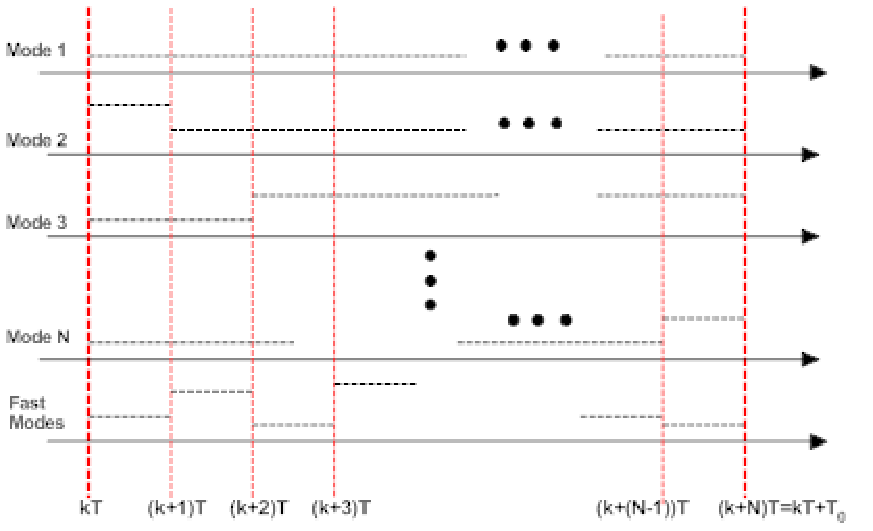}
\caption {Interlacing. Basic Procedure} \label{inter_basic}
\end{figure}

In the last years low-cost devices with low-power requirements are widely used. When consumer electronics such as Hard Disk Drives need be controlled or micro-boards such as Arduino, Raspberry Pi and so on are utilized in control automatic field, they need a proper control algorithm implementation. It is what is called a Resource Constrained Environment with limited processing capacity that requires a reduction of computational complexity.
The controllers designed with techniques such as $\mathcal{H}^\infty$ or QFT may lead to high order transfer functions with a wide dynamics range. It is obvious because there are different frequency zones for different design specifications.
Basically, the interlacing technique tries to decompose the single-rate controller's modes into fast and slow parts according to certain rules. This controller is supposed to be the ``fast controller". The idea is to apply the slow modes with slow frequency but with different input-output sampling strategies (for instance with a phase difference of a fast sampling period among each of them) in order to reduce the computational load in the slow frequency and make it uniform in a frame period. This basic idea is depicted in figure \ref{inter_basic}. 
Some contributions were introduced assuming this topic. In \cite{wu2004performance,ding2006multirate,wu2008precision} the controller interlacing implementation for track-following control for hard disk drive is considered. The performance and aliasing analysis is developed. In \cite{bhattacharya2009control} a LTI continuous controller is transformed into a multirate controller by discretization of slow and fast modes at different sampling periods (multiples by a factor of $N$ between them). Some different structures for interlacing the slow modes are studied using singular value decomposition. There are other trend using the interlacing for design step planning. In this context \cite{jia2008new} and \cite{lopez2010two} consider interlacing for obtaining the slow part regarding diverse performance criteria. In the first case, the design procedure consists on matching the fast rate controller output at the slow sampling instances and in the second case a $\mathcal{H}^\infty$ redesign is considered in the slow part. These last contributions were introduced trying to add some design step to a pure interlacing application without considering any specification.\\
In the current contribution, the starting point is the fast single-rate controller. The different interlacing structures analyzed must consider that the slow parts lead to an acceptable performance. A new treatment of real and complex controller's poles will be introduced. A complete frequency response study will be assumed for different blocks order implementation trying to establish some general rule. For this purpose a discrete lifting modeling for every input-output sampling strategies considered in interlacing implementation is introduced.
In next section, the procedure to transform the fast modes into slow ones is explained. Then some ideas about the discrete lifting procedure and the frequency response associated are considered. Afterwards it is discussed the different structures that can be assumed applying interlacing technique. The following section is devoted to examples. Some conclusions finish the exposition. 

\section{Interlacing Statement: Slow Modes}
\subsection{Preliminaries}
First of all some basic multirate definitions, operations and properties are needed.
$F^T$ will denote either the Z-transform of the sequence $\{ f(kT) \}$ obtained by sampling the continuous signal $f(t)$ or the sampling rate transformation of a discrete signal $F$. More explanation will be given below.
$$ F^T(z)= \mathcal{Z} ^T [ \{f(kT)\} ] =\sum_{k=0}^{\infty} f(kT) z^{-k} $$
in the same way, if the sampling period is $NT$
$$ F^{NT}(z^N)=\mathcal{Z} ^{NT}[ \{ f(kT) \} ]=\sum_{k=0}^{\infty} f(kNT) z^{-kN} $$
Now it is defined the upsampling (from $NT$ to $T$) and downsampling (from $T$ to $NT$) transforms:
$$ [F^{NT}(z^N)]^T=\bar{F}^T(z^N)=\sum_{k=0}^\infty \bar{f}(kT) z^{-kN} =
\begin{cases}  \bar{f}(kT)=f(kT); \hspace{0.3cm} \forall k=\lambda N \\
\bar{f}(kT)=0; \hspace{0.3cm} \forall k \neq \lambda N 
\end{cases} $$

$$ [F^{T}(z)]^{NT}=\hat{Y}^{NT}(z^N)=\sum_{k=0}^\infty f(kNT) z^{-kN}=F^{NT}(z^N) $$

Some upsampling-downsamling properties are:
\begin{equation*}
\begin{split} 
[X^T Y^T]^{NT} & \neq [X^T]^{NT}[Y^T]^{NT} \\
[X^{NT} Y^{NT}]^{T} & = [X^{NT}]^{T}[Y^{NT}]^{T}\\
[X^T[Y^{NT}]^T]^{NT} & =[X^T]^{NT}Y^{NT}
\end{split}
\end{equation*}

\subsection{Problem Statement}
Given a continuous controller $C(s)$ the Interlacing procedure is based on skipping slow and fast poles and applying them in different periods in the discretized version $C(z)$. The decomposition can be into series or parallel terms. Anyhow, the goal is to implement the slow part terms interspersed in such a way that a slow part fraction expansion will be needed. Therefore, when a single-rate controller is desired to implement with interlacing, some poles must be with different dynamics making sense to this operation. The first problem that is found is how to resample a fast sampling period slow pole to a slow sampling period slow pole.
Considering some dynamic areas selected according to some rule, a first decomposition is assumed:
\begin{equation*} 
\begin{split} 
C(s)=C_f(s)C_s(s) & =\frac{N_c(s)}{D_c(s)}=\frac{N_{cf}(s)}{D_{cf}(s)}\frac{N_{cs}(s)}{D_{cs}(s)}= \\
 &= \frac{K \prod_{i=1}^m (s+z_i) \prod_{p=1}^q (s^2+2\delta_p w_{np}+w_{np}^2)}{s^h \prod_{j=1}^m (s+p_j) \prod_{r=1}^s (s^2+2\delta_p w_{nr}+w_{nr}^2)} 
\end{split}
\end{equation*}
with $C_f(s)$ and $C_s(s)$ concerning fast and slow modes respectively (nothing it is said about the zeroes at the moment). Without loss of generality the series decomposition will be first discussed. The objective is to decompose the fast discrete controller slow part in parallel terms thinking in an interlacing implementation.
$$C^T_s(z)=\frac{N^T_{cs}(z)}{D^T_{cs}(z)}  $$
where $T$ indicates the discrete transfer function sampling period and $N^T_{cs}(z)$ and $D^T_{cs}(z)$ are polynomials in $z$, where the variable $z$ stands for the LTI z-transform argument at sampling period T, and consequently $z^N$ is related to $NT$. The poles and zeroes of $C^T_s(z)$ are denoted $\alpha_i$ and $\beta_j$ respectively.
Following \cite{lu1989least} is possible to transform a fast discrete transfer function into a form from which is viable to apply a fast input but every $N$ instants. Specifically, if a polynomial $W^T(z)$ is assumed:
$$W^T(z)=\displaystyle\prod_{i=1}^{n} \left( z^{N-1}+\alpha_{i,T}z^{N-2}+\hdots +\alpha_{i,T}^{N-1}\right)$$
then:
$$C^T_s(z)=\frac{U^T(z)}{E^T(z)}=\frac{N^T_{cs}(z)}{D^T_{cs}(z)}= \frac{N^T_{cs}(z)W^T(z)}{D^T_{cs}(z) W^T(z)}= \frac{\tilde{N}^T_{cs}(z)}{[D^{NT}_{cs}(z^N)]^T} = \frac{U^T(z)}{[E^{NT}(z)]^T} $$
being $E^T(z)$ the input and $U^T(z)$ the output. If it is supposed that the order of the denominator is greater or equal to the order of the numerator, the difference equation will be (case greater with second order denominator):
$$
y(kT)=b_1 u(kT-1)+b_2 u(kT-2)+b_{N-1} u(kT-(N-1)T)-a_1 e(kT-NT)-a_2 e(kT-2NT)
$$
that is, the input can be a $NT$ period signal and the output will be a $T$ sampling period signal.
If a new polynomial is defined:
$$ W_H^T=\left[ \frac{1-e^{NTs}}{s} \right]^T=\frac{1-z^{-N}}{1-z^{-1}}= 1+z^{-1}+\hdots +z^{N-1} $$
then it is possible to express:
$$
\left[ \frac{U^T(z)}{[E^{NT}(z)]^T} \right]^{NT}=\frac{[U^T(z)]^{NT}}{[[E^{NT}(z)]^T]^{NT}}=\frac{[U^T(z)]^{NT}}{E^{NT}(z)}=\frac{[ W_H^T \tilde{N^T_{cs}(z)}^T]^{NT}}{[[D^{NT}_{cs}(z^N)]^T]^{NT}}=\frac{N^{NT}_{cs}(z^N)}{D^{NT}_{cs}(z^N)}
$$
that is, a pure slow block, with slow input and output is obtained.

\subsection{Interlacing Implementation Strategies}
The following point is to discuss how can be implemented the slow part of the fast controller. It must be distinguished strategies for input and for output \cite{bhattacharya2009control}.
Regarding the input, the slow terms could assume:
\begin{itemize}
\item I-1. Fast Input
\item I-2. Slow Input
\item I-3 Mean Input 
\end{itemize}
In the first case, it is used the current fast input in each slow block of fast controller according to its order implementation (see figure \ref{inter_basic}). Note that each slow block update is occurring once every $N$ instants.
In the second case all slow blocks are fed by the same slow sampling valid in the $NT$ metaperiod. The third option is to consider the mean of the current and $(N-1)$ fast sampling of the input signal $E^T$ for the slow block input. This last option will not be considered because it is needed some additions.
\\
With respect to the output, it can be observed basically two options. The slow block output is kept during the $NT$ metaperiod and updated when the block is switched on (O-1) or every slow block output is stored and just all slow blocks addition is injected at the end of the metaperiod according to slow sampling times (O-2). O-1 will be noted like fast change or only fast and O-2 like slow change or only slow.
Some combinations of the different options will be analyzed with legends I-{1,2} and O-{1,2} for the input and output alternatives described before.
 
\section{Discrete Lifting Revisited}
A well-know method for modeling multirate systems is the so called Lifting. From the Kranc's idea (Vector Switch Decomposition) \cite{kranc1957input}, lifting is based on lifted the signals in one frame. That frame is the period at which every sampling sequence is repeated (metaperiod). In general, for a Dual-Rate system with input and output sampling periods $T_u$ and $T_y$ (rationally related) respectively, $T_0=lcm(T_u,T_y)$ will be the metaperiod and the input and output signals will be lifted in vectors with size $N_u$ and $N_y$ respectively, such that $T_0=T_uN_u=T_yN_y$. In the case of this contribution, schemes with $T$ and $NT$ samplers will be considered. Therefore, the metaperiod will be $T_0=NT$ and the discrete $T$ signals will be lifted in size $N$ vectors.
It is usual to model the behaviour of the Dual-Rate system characterized via a ``lifted" transfer function matrix:
\begin{equation}
y_l(z^N) = \tilde{G}(z^N)u_l(z^N) (4)
\end{equation}
where the subindex "`l"' denotes "`lifted"' and $z^N$ is referred to the $z$ variable at least common period $T_0$. It is possible to assume internal or external representations for lifting models; in \cite{francis1988stability} the links between them are described. In this paper, the internal representation is the proper one. In order to explain how is deduced, a general strictly proper  continuous system preceded by an ZOH and samplings period $T_u$ for input and $T_y$ for output will be considered. It is possible to obtain the $T_0$ lifting representation from a $T$ ZOH-discretization of the process being $T=gcd(T_u,T_y)$. If that discretization is $(A,B,C,0)$, the lifting model can be deduced by repeated evaluations of the equations at sampling period $T$. The result will be:   
\begin{equation} \label{x_con_zoh}
\begin{split}
y(kT_0 +\zeta T)=&Cx(kT_0 + \zeta T)=\\
&=C [ A^{\zeta} x(kT_0)+ A^{\zeta -1} Bu(kT_0)+ \\
 &+A^{\zeta -2} Bu(kT_0+T)+\hdots + Bu(kT_0+(\zeta -1)T)  ] 
\end{split}
\end{equation}
for $\zeta=1,\dots,(N_u -1) N_y$,. However, the zero-order-hold
entails
\begin{equation} \label{u_con_zoh}
\begin{split}
&u(kT_0+d N_y T)=u(kT_0+(dN_y+1)T)=\dots=u[kT_0+((d+1)N_y -1) T]\\
& \forall   d=0, 1 \dots, (N_u-1)
 \end{split}
\end{equation}
The lifted matrices $(A-l,B_l,C_l,D_l)$ are obtained by suitably stacking the results from the above equations.\\
For instance, the lifted model in the case with $T_u=T$ and $T_y=T_0==4T$ will be:
\begin{equation}\label{eqlfm}
\begin{array}{lll}
x[(k+1)T_0]&=&A^4x(kT_0)+\left(\begin{array}{cccc} A^3 B & A^2 B & A B & B ) \end{array} \right) \left(\begin{array}{c} u(kT_0) \\ u(kT_0+T)\\ u(kT_0+2T) \\  u(kT_0+3T) \end{array} \right) \\
 
\left(\begin{array}{c} y(kT_0)  \end{array} \right) &= &\left(\begin{array}{c} C \end{array}\right)x(kT_0)+\left(\begin{array}{cccc} CA^3B & CA^2B &CAB &CB  \end{array}\right) \left(\begin{array}{c} u(kT_0) \\ u(kT_0+T)\\ u(kT_0+2T) \\  u(kT_0+3T) \end{array} \right)
\end{array}
\end{equation}

\section{Interlacing Lifting Application} \label{sec_block}
It is clear that the successive switches of all slow blocks leads to a periodic operation which needs the lifting procedure in order to obtain a LTI function.
In this case a special care must be considered when the discrete lifting modeling is applied on this environment. First of all the complete operation is $NT$ periodic. So, if lifting is wanted to be applied, the blocks order implementation should be studied in one metaperiod. There will be different modeling options which are going to be analyzed. The problem with $N$ slow poles at fast controller is considered. For all input cases I-{1,2} the problem is similar because a certain signal value arrives at the moment of the switch of a certain slow block. After the current explanation some considerations will be made about the input signal selection. Now, the problem is that for blocks $i=2 \hdots N$ there are $(i-1)T$ sampling periods where the output signal is the same that at preceding metaperiod and $NT-(i-1)T$ outputs computed with the current input signal value. For the first control actions the lifting modeling requires a new state which will be a dummy variable $\chi$ that will represent the control action at the end of the previous metaperiod.
\begin{equation}
\left(
\begin{array}{c}
x\\
\chi \\
\end{array}
\right) _{(k+1)T} =
\left( \begin{array}{lc}
A_i & B_i \\
0 & 0 \\  
\end{array} 
\right) 
\left(
\begin{array}{c}
x\\
\chi \\
\end{array}
\right) _{kT} +
\left(
\begin{array}{c}
0\\
1\\
\end{array}
\right) e_k
\end{equation}
\begin{equation}
u_k = 
\left( \begin{array}{lc}
C_i & D_i \\
\end{array} 
\right) 
\left(
\begin{array}{c}
x\\
\chi \\
\end{array}
\right) _{kT} +
\left(
\begin{array}{c}
0\\
\end{array}
\right) e_k
\end{equation}
being $(A_i,B_i,C_i,D_i)$ the state space matrices of slow block $i$ at sampling period $NT$.
Note that just one dummy variable is needed because the control action follows from just one previous metaperiod.
Usually the lifting procedure uses a quadruple representation packing the matrices from a general state space representation $\textit{A},\textit{B},\textit{C},\textit{D}$:
\begin{equation}
Block_i \equiv \left(
\begin{array}{c|c}
\textit{A} & \textit{B} \\ \hline
\textit{C} & \textit{D} \\
\end{array}
\right)
\end{equation}
As it was said before, the input signal treatment completes the lifting modeling.
The expression for representing the fast input selection into a metaperiod will be:
\begin{equation}
Block_i 
\left(
\begin{array}{c}
0\\
\vdots \\
1 \hspace{0.2cm} \text{file i}\\
\vdots \\
0 \\
\end{array}
\right) ^{t}
\end{equation}
that is with a ``1" selecting the input instant order in the metaperiod. Note that in one metaperiod $NT$ there are $N$ values lifted from fast signals at $T$.
Finally the sampling period updating in the metaperiod will be described by means of a similar operation. For instance for block $i=3$ the contribution every fast sampling period into a metaperiod will be:
\begin{equation}
\left(
\begin{array}{c}
1\\
1\\
1\\
0\\
\vdots\\
0 \hspace{0.2cm} \text{file N}\\
\end{array}
\right) block_{\chi_{i}} +
\left(
\begin{array}{c}
0\\
0\\
0\\
1\\
\vdots \\
1 \hspace{0.2cm} \text{file N}\\
\end{array}
\right) block_{i}
\end{equation}
As an example to understand the procedure, it is considered the case of a fast controller with four modes $b_j$ being three of them $b_2,b_3,b_4$ slow ones. It is considered that the implementation is $b_1$ fast and $b_4,b_2,b_3$ with this order. Therefore $N=3$ and the option (I-1,O-1), will require the following open-loop lifting modeling:
%& [1 1 1]'*b4k*[1 0 0]+
%&+[1 0 0]'*b2k*[0 1 0]+[0 1 1]'*b2ak*[0 1 0]+[1 1 0]'*b3k*[0 0 1]+[0 0 1]'*b3ak*[0 0 1]+b1k;
\begin{equation}
\begin{split}
&\left( \begin{array}{ccc} 1 &1 &1 \end{array} \right)' *bk4* \left( \begin{array}{ccc} 1 &0 &0 \end{array} \right)+ \\
&+ \left[ \left( \begin{array}{ccc} 1 &0 &0 \end{array} \right)'*bk2+\left( \begin{array}{ccc} 0 &1 &1 \end{array} \right)'bk2_{\chi}  \right]*\left( \begin{array}{ccc} 0 &1 &0 \end{array} \right)+ \\
&+ \left[ \left( \begin{array}{ccc} 1 &1 &0 \end{array} \right)'*bk3+\left( \begin{array}{ccc} 0 &0 &1 \end{array} \right)' bk3_{\chi}  \right]*\left( \begin{array}{ccc} 0 &0 &1 \end{array} \right) \\
& +bk1
\end{split}
\end{equation}
The block $bk1$ does not need the selector vectors because is used in all fast instants.\\
The case (I-1,O-2) will be modeled by:
\begin{equation}
\begin{split}
&\left( \begin{array}{ccc} 1 &1 &1 \end{array} \right)' *bk4* \left( \begin{array}{ccc} 1 &0 &0 \end{array} \right)+ \\
&+ \left[ \left( \begin{array}{ccc} 1 &0 &0 \end{array} \right)'*bk2+\left( \begin{array}{ccc} 0 &1 &1 \end{array} \right)'bk2_{\chi}  \right]*\left( \begin{array}{ccc} 1 &0 &0 \end{array} \right)+ \\
&+ \left[ \left( \begin{array}{ccc} 1 &1 &0 \end{array} \right)'*bk3+\left( \begin{array}{ccc} 0 &0 &1 \end{array} \right)' bk3_{\chi}  \right]*\left( \begin{array}{ccc} 1 &0 &0 \end{array} \right) \\
& +bk1
\end{split}
\end{equation}

\section{Dual-Rate Frequency Response} \label{drfr}
When a Dual-Rate system is studied in the frequency domain, it is usual to consider it like a multivariable system (see lifting modeling). A classical perspective is to analyze the singular value decomposition of the lifted matrix. The problem with such a procedure is that some information is lost. In \cite{salt2014new} an efficient algorithm for computing the Dual-Rate system frequency response was introduced. The result obtained in that contribution establish that the output $y(k)$, when $u(k)=e^{j\omega T_u k}$, of a SISO
dual-rate ($N_uT_u=N_yT_y$) lifted system
$y_{l}(z)=G_{lifted}(z)u_l(z)$ is a collection of components
$y_r(k)=\bar{y}_r e^{j T_y\omega_rk}$ of frequencies
$\omega_r=\omega +2\omega^s_y r/N_y$, for $r=0,\dots,(N_y-1)$, with
$\omega^s_y=\pi/T_y$, and $\bar{y}_r$ is given by:
\begin{equation}\label{eqt3}
\bar{y}_r=\frac{1}{N_y}\sum_{p=0}^{N_y-1}\sum_{q=0}^{N_u-1}
G_{pq}(e^{j\omega_r T_y N_y})e^{-j T_y\omega_r p}e^{j\omega T_uq}
\end{equation}
It is possible to check that,  from \eqref{eqt3}, the components will be given by the product
of the frequency response of a left factor:
$$[1 \ z^{-1} \ z^{-2} \ \dots \ z^{-(N_y-1)}]G_{lifted}(z^{N_y})$$
replacing $z=e^{j\omega_rT_y}$, which gives a row vector, and the
right factor (column vector)
$$(1\ z \ z^2 \ \dots z^{N_u-1})^T $$
replacing $z=e^{j\omega T_u}$. 
If it was stated before, $T_u=T$ and $T_y=T_0=NT$, there will be $N_y$ components which can be read from only one Bode plot from the lifted matrix \footnote{It is possible to obtain the frequency response either in internal or the equivalent external representation}. For an input frequency $d$ Rad/s, the readings must be done at $d,d+w_s,\hdots d+(N-1)w_s$ being $w_s=\frac{2\pi}{NT}$.\\
This result will be assumed to detect the existence of ripple applying interlacing implementation, but as it is easy to understand does not lead to pure frequency response, because the "$N$ sinusoids addition is not a sinusoid. For that purposes the metaperiod sum must be considered:
\begin{equation*}
\begin{split}
 y(kT_y)=&A_1 sin(wkT_y+\varphi_1)+A_2 sin((w+N_u w_s)kT_y+\varphi_2)+\hdots+\\
&+ A_{N_y} sin((w+(N_y -1)N_u w_s)kT_y+\varphi_{N_y})
\end{split}
\end{equation*}
As it was said this is not a pure sine signal. However if the decomposition of each component is observed:
%\begin{equation*}
%\begin{split}
%A_\nu sin (w_\nu k T_y + \varphi_{\nu})=&A_\nu sin (w_\nu k T_0 + \varphi_{\nu})+\\
%&z ^{-1}_{T_y} A_\nu sin (w_\nu (k T_0+T_y) + \varphi_{\nu})+\hdots\\
%&z ^{-(N_y -1)}_{T_y} A_\nu sin (w_\nu (k T_0+(N_y -1)T_y) + \varphi_{\nu})+\\
%& for \hspace{0.3 cm} \nu=1,\hdots, (N_y -1) \\
%\end{split}
%\end{equation*}
%\\
\begin{equation*}
\begin{split}
A_\nu sin (w_\nu k T_y + \varphi_{\nu})=&A_\nu sin (w_\nu k T_0 + \varphi_{\nu})+\\
&z ^{-1}_{T_y} A_\nu sin (w_\nu k T_0+ (w_\nu T_y  + \varphi_{\nu}))+\hdots\\
&z ^{-(N_y -1)}_{T_y} A_\nu sin (w_\nu k T_0+(w_\nu (N_y -1)T_y + \varphi_{\nu}))+\\
& for \hspace{0.3 cm} \nu=1,\hdots, (N_y -1) \\
\end{split}
\end{equation*}
\\
So, adding the contributions of each component at $kT_0$, that is the downsampling of $y(kT_y)$:
\begin{equation*}
\begin{split}
y^{T_0}(kT_y)&=A_1 sin (w k T_0 + \varphi_{1})+\\
&+A_2 sin (w+w_s) k T_0+ ((w_+w_s) T_y  + \varphi_{\nu}))+\hdots \\
&+A_{N_y} sin(((w+(N_y -1) w_s)+(N_y -1)N_u w_s)kT_0+\varphi_{N_y})\\
\end{split}
\end{equation*}
That finally drives to the $T_0$ approximation:
\begin{equation*}
\begin{split}
y^{T_0}(kT_y)&=A_1 sin (w k T_0 + \varphi_{1})+\\
&+A_2 sin (w k T_0+ (w_s T_0  + \varphi_{\nu}))+\hdots \\
&+A_{N_y} sin((w kT_0+ ((N_y -1) w_s T_0 + \varphi_{N_y}))\\
\end{split}
\end{equation*}
that is a pure sinusoid. In this process, the detail is lost but allows to consider the classical frequency representation.
However, in this paper, either the components and the $T_0$ sum will be assumed. The components lead to a suitable mean to identify ripple in the blocks contributions; actually this is a common problem in multirate systems if the sampling periods are not appropriate.

\section{Applications}
\begin{figure} \centering
\includegraphics[height=8cm, width=10cm]{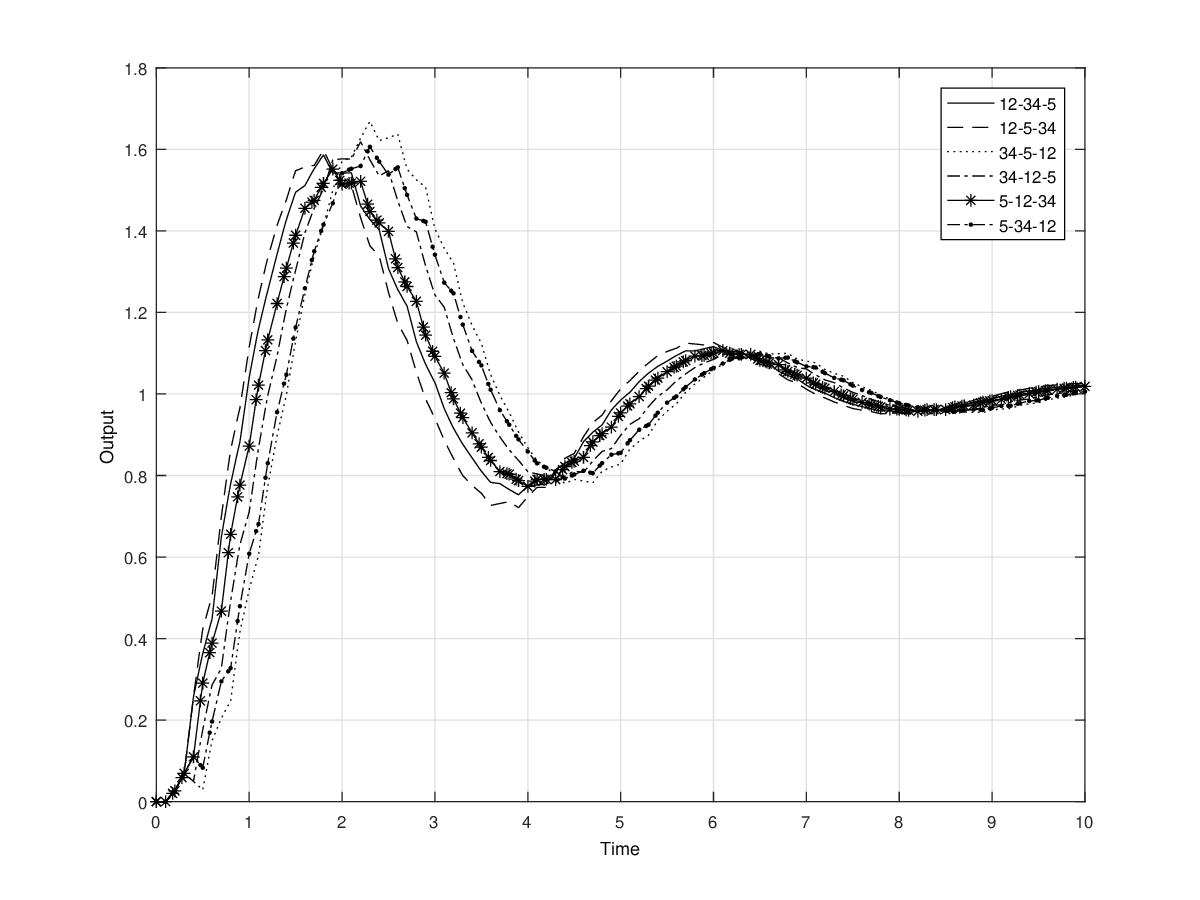}
\caption {Closed Loop Output Response with different Blocks order. Fast Input-Fast output} \label{blockorder}
\end{figure}

\begin{figure} \centering
\includegraphics[height=8cm, width=10cm]{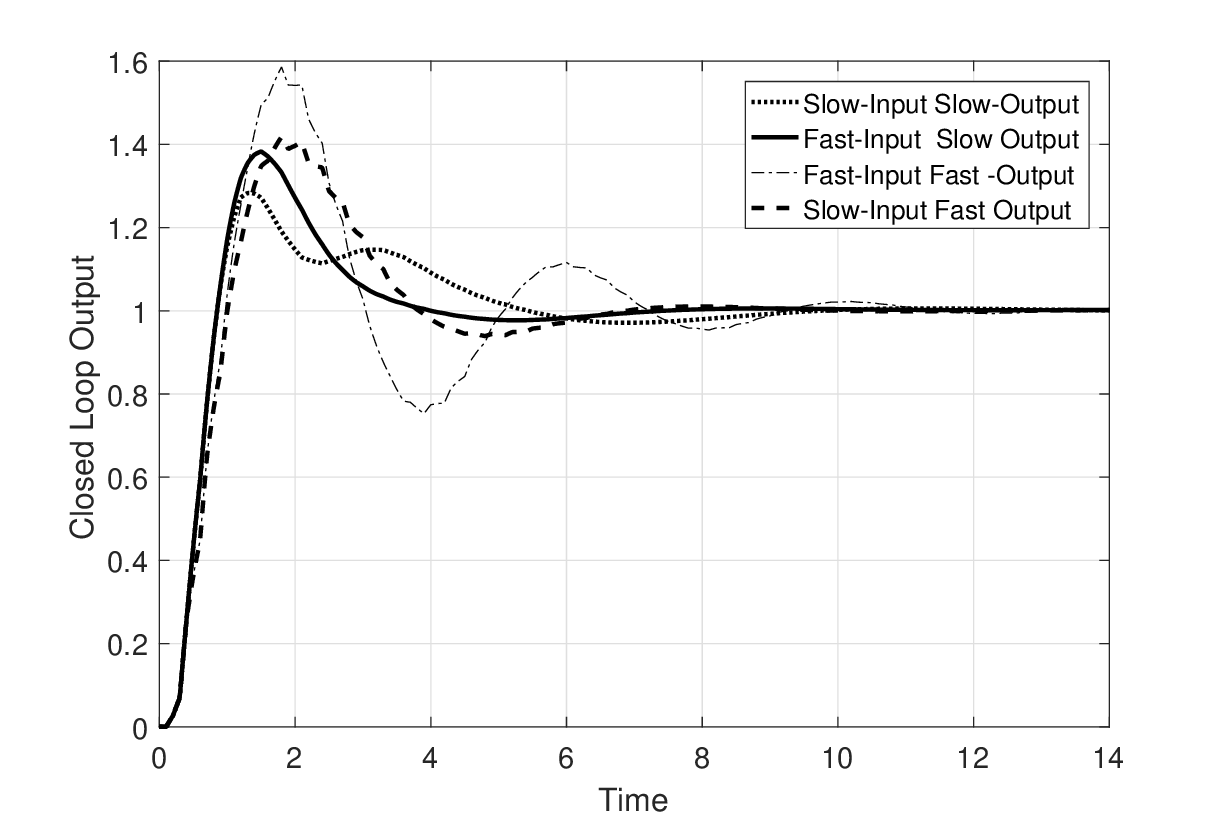}
\caption {Closed Loop Output Response with different Strategies Input - Output} \label{Inp_Out_dif}
\end{figure}

\begin{figure} \centering
\includegraphics[height=8cm, width=10cm]{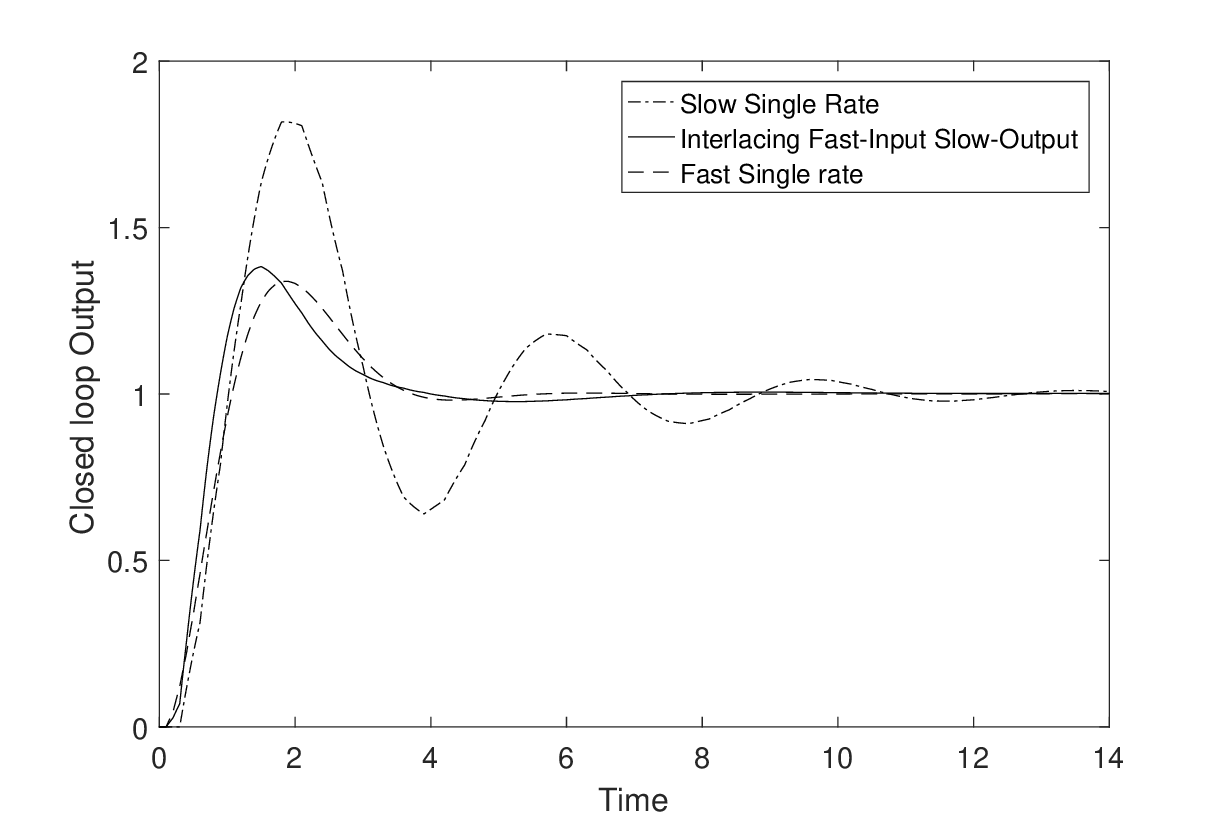}
\caption {Closed Loop Output Response Comparison with Single Rate Controllers} \label{SR_comparison}
\end{figure}

Now, a specific problem is considered. A controller is designed for the process:
$$
G(s)=\frac{(s+1)}{s^2+2s+1.5}
$$
The control design procedure is a mixed sensitivity $\mathcal{H}^\infty$ based on. The continuous controller obtained with this technique was:
$$
C(s)=\frac{5.0295 (s+0.6524) (s+0.05)^3 (s+0.02857) (s^2 + 2s + 1.5)}{(s+3.488) (s+0.9743) (s+0.04)^2 (s+0.02727) (s+0.0024) (s^2 + 0.1295s + 0.004509)}    
$$
A proper sampling period (fast) is $T=0.1$. The discretization leads to the following minimum realization discrete controller:
$$
C_d(z)=\frac{0.46177 (z-0.9976) (z-0.9937) (z-0.9368) (z^2 - 1.806z + 0.8191)}{(z-1)^2 (z-0.9072) (z-0.7056) (z^2 - 1.973z + 0.9732)}
$$
with poles $1 \hspace{0.1cm} \text{Double}, 0.9865 \pm 0.0055i, 0.9072, 0.7056 $.
As it can be seen, applying a mild dynamics separation rule, there is just one fast pole. Therefore, a series configuration will be:
\begin{equation*}
\begin{split}
C_d & =C_{df} C_{ds} \\
C_{df} &=\frac{N_{df}}{z-0.7056} \\
C_{ds} &=\frac{N_{ds}}{(z-1)^2 (z-0.9072) (z^2 - 1.973z + 0.9732)}
\end{split}
\end{equation*}
and it is possible to follow some rules for the numerator groups separation. However, if a parallel decomposition is considered:
\begin{equation*}
\begin{split}
C_d & =C_{df}+ C_{ds} \\
C_{df} &=\frac{0.2798}{z-0.7056} \\
C_{ds} &=\frac{0.4793 z - 0.4782}{(z-1)^2 } + \frac{ -0.2567 z + 0.2591}{z^2 - 1.973z + 0.9732} +\frac{ -0.03991}{z-0.9072}
\end{split}
\end{equation*}
Note that in the last case no decision is needed about the numerators separation.
As it was said, there are two blocks with a special treatment. They are the related ones to the double pole and the complex pair poles which treatment can be the same.
In both cases, the use of polynomial $W^T$ allows to obtain the modified $T$ fast terms:
\begin{equation*}
\begin{split}
C_{s,12}^T & = \frac{0.4793 z^5 + 0.4803 z^4 + 0.4814 z^3 - 0.4762 z^2 - 0.4772 z - 0.4782}{z^6 - 2 z^3 + 1} \\
C_{s,34}^T & =\frac{-0.2567 z^5 - 0.2474 z^4 - 0.2382 z^3 + 0.2636 z^2 + 0.2544 z + 0.2454}{z^6 - 1.92 z^3 + 0.9218} \\
C_{s,5}^T & =\frac{-0.03991 z^2 - 0.0362 z - 0.03284}{z^3 - 0.7465}
\end{split}
\end{equation*}
where the subindex $(i(,j))$ describes the number of poles and an arbitrary order of the blocks.
After using the polynomial $W_R$, and a downsampling operation the transformation into a slow term is completed:
\begin{equation*}
\begin{split}
C_{s,12}^{NT} &=\frac{1.441 z - 1.432}{z^2 - 2 z + 1} \\
C_{s,34}^{NT} &=\frac{-0.7423 z + 0.7634}{z^2 - 1.92 z + 0.9218} \\
C_{s,5}^{NT} & =\frac{-0.109}{z - 0.7465}
\end{split}
\end{equation*}
Note that the blocks $C_{s,i}^T$ corresponds to the $block_i$ in section \ref{sec_block}.
Before the interlacing can be applied, it is necessary to test the slow single rate controller. In the case that this slow control would result in losing performance or even instability then it would be inadvisable to consider this kind of implementation.
The closed loop output time response with different interlacing blocks order in the case (I-1,O-1) is depicted in figure \ref{blockorder}. Different responses are obtained based on the different gains of the different blocks. It is observed that the response has a light chattering effect due to the fast switch of slow interlacing blocks.\\
Figure \ref{Inp_Out_dif} shows the different closed loop outputs with different Input-Output implementation strategies. It seems that the fast input sampling and slow updating of slow interlacing blocks leads to a better response. The comparison of this last curve (I-1, O-1) with single rate (slow and fast) controllers is represented in figure \ref{SR_comparison}. In general, the responses using fast controller interlacing should be between both single rate (slow and fast) controllers. That is the reason why is important that the slow single rate response was suitable. Obviously the ideal solution would be a good approximation of the interlacing problem to the fast single rate case. 

\begin{figure} \centering
\includegraphics[height=9cm, width=12cm]{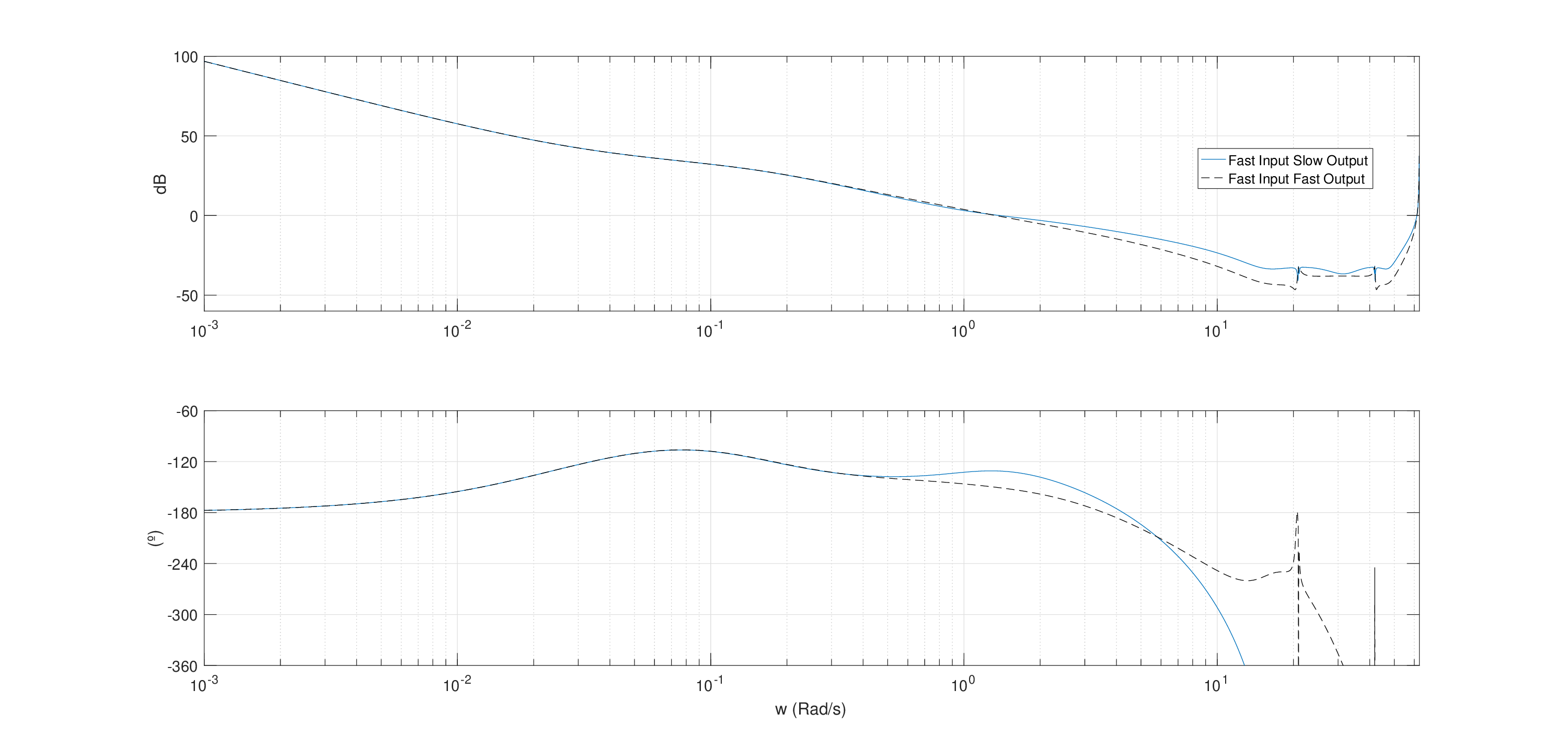}
\caption {Dual-Rate Bode diagrams with different Input - Output strategies} \label{Fast_bodes}
\end{figure}

Using the multirate frequency response tool by author introduced in section \ref{drfr} is possible to analyze some interesting characteristics of this kind of implementation. In figure \ref{Fast_bodes} are showed the dual-rate Bode diagrams assuming different input-output strategies. The high slope around $20$ Rad/s explains the chattering showed in figure \ref{blockorder}.
The frequency response in figure \ref{Fast_bodes} obtained from lifting representation of both cases (I-1,O-1) and (I-1, O-2) validates the time simulation with better phase margin ($50^{\circ}$ vs $30^{\circ}$) in the case (I-1, O-2), that is with fast input and slow output. But this case has a slightly less gain margin.
\begin{figure} \centering
\includegraphics[height=9cm, width=12cm]{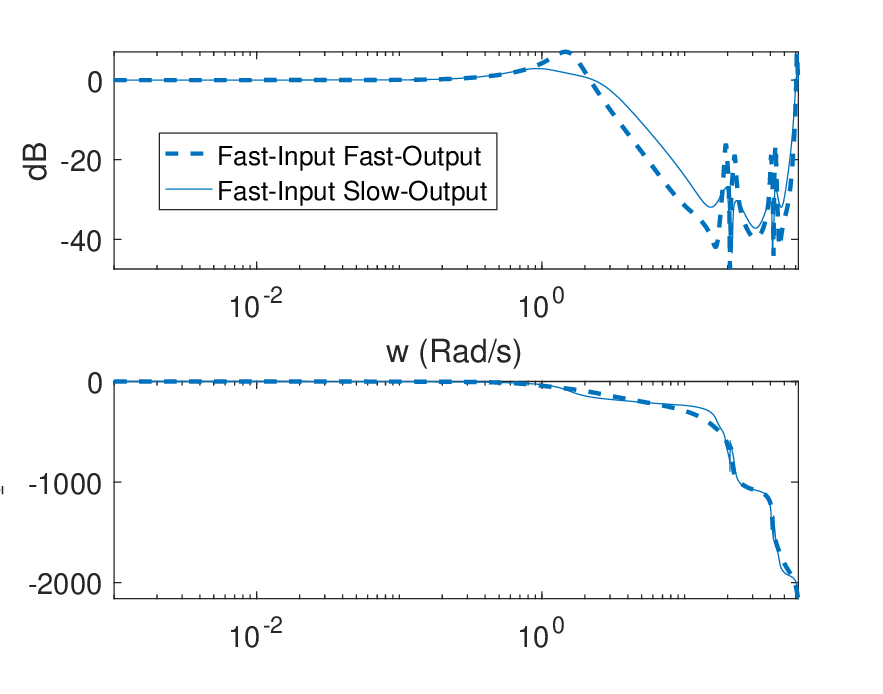}
\caption {Close Loop Dual-Rate Bodes with different Strategies Input - Output} \label{CLoop_bodes}
\end{figure}
The dual-rate Bode diagrams of closed loops in figure \ref{CLoop_bodes} reveals that the (I-1,O-1) exhibits better bandwidth as it was known by time responses, but the case (I-1,O-2) has a better sensitivity to high frequency noise.
\section{Conclusions}
This paper has presented a clear and systematic procedure for obtaining the model of a controller implemented using the interlacing technique, which results in an LPV system.  This modeling, based on techniques specific to multi-frequency control, allows the new properties of the controller implemented in this way to be analyzed. By taking advantage of an algorithm specific to the frequency response of this type of system, conclusions can be drawn about different possible configurations of this type of implementation. A future line of research would be to study the order of application of the different elements that make up the decomposition to be considered in interlacing.

\section*{Agradecimientos}

Este trabajo se ha realizado gracias al apoyo del Proyecto PID2023-151755OB-I00, financiado por MICIU/AEI/10.13039/501100011033/ y por FEDER/UE.
%
%\section{References}
\bibliographystyle{elsarticle-num}

\bibliography{bookmv,AV_multirate}

\begin{thebibliography}{10}
\expandafter\ifx\csname url\endcsname\relax
  \def\url#1{\texttt{#1}}\fi
\expandafter\ifx\csname urlprefix\endcsname\relax\def\urlprefix{URL }\fi
\expandafter\ifx\csname href\endcsname\relax
  \def\href#1#2{#2} \def\path#1{#1}\fi

\bibitem{wu2004performance}
S.~Wu, M.~Tomizuka, Performance and aliasing analysis of multi-rate digital
  controllers with interlacing, in: American Control Conference, 2004.
  Proceedings of the 2004, Vol.~4, IEEE, 2004, pp. 3514--3519.

\bibitem{ding2006multirate}
J.~Ding, F.~Marcassa, S.~Wu, M.~Tomizuka, Multirate control for computation
  saving, Control Systems Technology, IEEE Transactions on 14~(1) (2006)
  165--169.

\bibitem{wu2008precision}
S.-C. Wu, Precision control for high-density and cost-effective hard disk
  drives, University of California, Berkeley, 2008.

\bibitem{bhattacharya2009control}
R.~Bhattacharya, G.~Balas, Control in computationally constrained environments,
  IEEE transactions on control systems technology 17~(3) (2009) 589--599.

\bibitem{jia2008new}
Q.~Jia, A new method of multirate state feedback control with application to an
  hdd servo system, Mechatronics 18~(1) (2008) 13--20.

\bibitem{lopez2010two}
S.~L{\'o}pez-L{\'o}pez, A.~Sideris, J.~Yu, Two-stage design of multirate h8
  optimal controllers, in: American Control Conference (ACC), 2010, IEEE, 2010,
  pp. 2647--2652.

\bibitem{lu1989least}
W.~Lu, D.~G. Fisher, Least-squares output estimation with multirate sampling,
  IEEE Transactions on Automatic Control 34~(6) (1989) 669--672.

\bibitem{kranc1957input}
G.~Kranc, Input-output analysis of multirate feedback systems, Automatic
  Control, IRE Transactions on 3~(1) (1957) 21--28.

\bibitem{francis1988stability}
B.~Francis, T.~Georgiou, Stability theory for linear time-invariant plants with
  periodic digital controllers, Automatic Control, IEEE Transactions on 33~(9)
  (1988) 820--832.

\bibitem{salt2014new}
J.~Salt, A.~Sala, A new algorithm for dual-rate systems frequency response
  computation in discrete control systems, Applied Mathematical Modelling
  38~(23) (2014) 5692--5704.

\end{thebibliography}

\end{document}